\documentclass[pss,fleqn,embeddedheads]{w-art}
\usepackage{times}
\usepackage{w-thm}
\numberwithin{equation}{section}
\usepackage[]{graphicx}
\graphicspath{{.}{./EPS/}}
\begin{document}
%%    The information for the title page will be placed between
%%    \begin{document} and \maketitle. The order of most entries
%%    is determined by the class file and can not be changed by
%%    rearranging them. The maketitle command follows after the
%%    abstract.
%%
%%    Most of the following commands will be completed by the publisher.
%%
%%    The copyrightyear is defined in the .clo file as the first argument
%%    of the copyrightinfo command. If the copyrightyear differs from that
%%    value it might be adjusted by the following definition:
%%
\renewcommand{\copyrightyear}{2006}% uncomment to change the copyrightyear.
\DOIsuffix{theDOIsuffix}
%%
%% issueinfo for header and copyright line
\Volume{VV}
\Issue{I}
\Month{MM}
\Year{2006}
%%
%%    First and last pagenumber of the article. If the option
%%    'autolastpage' is set (default) the second argument may be left empty.
\pagespan{1}{}
%%
%%    Dates will be filled in by the publisher. The 'reviseddate' and
%%    'dateposted' (Published online) entry may be left empty.
\Receiveddate{PASPS-IV paper number PB-45}
%\Reviseddate{}
%\Accepteddate{}
%\Dateposted{}
%%

%% \pretitle{Editor's Choice}

%% We have a short and a long form for the title. The short form
%% (optional argument) goes into the running head.

\title{Electronic and spin properties of hole point contacts}

%% Please do not enter footnotes or \inst{}-notes into the optional
%% argument of the author command. The optional argument will go into
%% the header.  If there is only one address the marker \inst{x} may be
%% omitted.

%% Information for the first author.
\author[U. Z\"ulicke]{U. Z\"ulicke\footnote{Corresponding
     author: e-mail: {\sf u.zuelicke@massey.ac.nz}, Phone: +64\,6\,350\,5799\,extn\,7259,
     Fax: +64\,6\,350\,5682}\inst{1}} \address[\inst{1}]{Institute of Fundamental Sciences 
     and MacDiarmid Institute for Advanced Materials and Nanotechnology, Massey 
     University, Private Bag 11~222, Palmerston North, New Zealand}
%%
%%    Information for the second author
%\author[Sh. Second Author]{L. Second Author\footnote{Second author footnote.}\inst{1,2}}
%\address[\inst{2}]{Second address}
%%
%%    Information for the third author
%\author[Sh. Third Author]{L. Third Author\footnote{Third author footnote.}\inst{2}}
%%
%%    \dedicatory{This is a dedicatory.}
\begin{abstract}
We have studied theoretically the effect of a tuneable lateral confinement on
two-dimensional hole systems realised in III-V semiconductor heterostructures. Based
on the $4\times 4$ Luttinger description of the valence band, we have calculated
quasi-onedimensional (quasi-1D) hole subband energies and anisotropic Land\'e
$g$-factors. Confinement-induced band mixing results in the possibility to manipulate
electronic and spin properties of quasi-1D hole states over a much wider range than is
typically possible for confined conduction-band electrons. Our results are relevant for
recent experiments where source-drain-bias spectroscopy was used to measure Zeeman
splitting of holes in p-type quantum point contacts.
\end{abstract}
%% maketitle must follow the abstract.
\maketitle                   % Produces the title.

%% If there is not enough space inside the running head
%% for all authors including the title you may provide
%% the leftmark in one of the following three forms:

%% \renewcommand{\leftmark}
%% {First Author: A Short Title}

%% \renewcommand{\leftmark}
%% {First Author and Second Author: A Short Title}

%% \renewcommand{\leftmark}
%% {First Author et al.: A Short Title}

%% \tableofcontents  % Produces the table of contents.

\section{Introduction}
Low-dimensional hole systems provide an interesting playground for engineering spin
properties of charge carriers. In typical semiconductors, quantum confinement
affects the physical properties of holes much more strongly than those of
electrons~\cite{bastardrev}. This is because subband quantisation causes an energy
splitting between heavy-hole (HH) and light-hole (LH) bands, and modifies their residual
coupling. The Zeeman splitting of 2D hole systems for in-plane magnetic fields is a good
example; it is suppressed for HH states~\cite{OldHoleZ1} but doubled for LH
states~\cite{landwehr:94}. On the most basic level, this behaviour can be understood by
noting that the heavy and light-hole bands belong to a quadruplet of states having total
angular momentum $j=3/2$. More detailed theory~\cite{rolandbook} based on the
Luttinger description~\cite{luttham2} of the valence band provides realistic values for the
Land\'e $g$-factors of 2D holes, including the in-plane Zeeman-splitting anisotropy in
low-symmetry heterostructures. The competition between confinement-induced
band mixing and HH-LH energy splitting also determines the physical properties of hole
quantum wires~\cite{sercel:apl:90}, dots~\cite{darn:prb:94}, and localised acceptor
states~\cite{haug:prl:06}. Recently, transport measurements in hole 
wires~\cite{picc:apl:05} and point contacts~\cite{romain:apl:06} have
become possible, opening up new possibilities to investigate electronic and spin
properties of quasi-1D hole systems. Here we investigate the
confinement dependence of hole $g$-factors in quasi-1D structures, mapping the
crossover between the weakly confined 2D and symmetrically confined 1D limits.

\section{Theoretical description of quasi-1D hole systems}
The HH and LH bands are distinguished by the quantum number $j_z$ of projection of
total angular momentum along the quantisation ($z$) axis:  $j_z=\pm 3/2$ for HHs,
$j_z=\pm 1/2$ for LHs. Their dynamics is described by the Luttinger
Hamiltonian~\cite{luttham2}, which forms the basis for our investigation of quasi-1D hole
systems. While the effect of remote bands (split-off, conduction,
etc.) may need to be included to achieve reasonable quantitative accuracy~\cite
{citrin:prb:89}, we do not expect it to change the qualitative features of the 2D-to-1D
crossover focussed on in this work. In the following, we use atomic units ($\hbar= m_0 =
1$) and adhere to the hole picture for energy bands. We also find it useful to employ a
universal representation~\cite{lip:prl:70,fish:prb:95} of the
Luttinger Hamiltonian as a sum over tensor invariants where all
information about band structure is contained in the Luttinger parameters~\cite{luttham2, 
fish:prb:95} $\gamma_1, \gamma_{\mathrm s}=(3 \gamma_3 + 2 \gamma_2)/5,
\gamma_\delta=(\gamma_3 - \gamma_2)/2$, and coordinate-system-dependent
constants $c_j$. The most general $4\times 4$ valence-band Hamiltonian~\cite
{rolandbook,fish:prb:95} also includes terms describing spin splitting due to the Zeeman
effect or arising from bulk (B) and/or structural (S) inversion asymmetry (IA). It reads
\begin{equation}
{\mathcal H}_{4\times 4} = \left[ \frac{\gamma_1}{2} + \frac{5}{4}\,\gamma_{\mathrm s}
\right] {\hat{\mathbf k}}^2 -\gamma_{\mathrm s} \left(\hat{\mathbf k} \cdot \hat{\mathbf J}
\right)^2 + \gamma_\delta\sum_j c_j {\mathcal T}_j(\hat{\mathbf k}, \hat{\mathbf J}) +
{\mathcal H}_{\mathrm Z} + {\mathcal H}_{\mathrm{BIA}} + {\mathcal H}_{\mathrm{SIA}}
+ V(\hat{\mathbf r}) \,\, .
\end{equation}
We denote operators of position, kinetic wave vector, and total angular momentum by
$\hat{\mathbf r}$, $\hat{\mathbf k}$, and $\hat{\mathbf J}$, respectively. ${\mathcal
T}_j(\hat{\mathbf k}, \hat{\mathbf J})$ are tensor invariants associated with axial ($j=0$)
and cubic ($j\ne 0$) corrections to the spherical Luttinger Hamiltonian; see the notation
used by Fishman~\cite{fish:prb:95}.

In the following, we include the isotropic Zeeman term ${\mathcal H}_{\text{Z}} = 2 \kappa
\mu_{\text{B}} \, {\mathbf B} \cdot {\hat{\mathbf J}}$ with bulk-hole $g$-factor $\kappa$ but
neglect spin splitting due to IA. We consider a rectangular hard-wall confining potential
$V(\hat{\mathbf r})$ in the $xy$ plane, characterised by quantum-well widths $W_x$ and
$W_y$. Hence we adopt the wire axis as the quantisation axis of total angular momentum.
As we are interested in describing the properties of hole point contacts realised by lateral
confinement in a 2D HH system~\cite{romain:apl:06}, we assume a strong confinement in
$x$ direction and consider only the lowest orbital quantum-well bound state. Results
obtained in the spherical approximation (i.e., for $\gamma_\delta=0$) serve to illustrate
that the crossover between the 2D and symmetrically confined 1D limits (realised for
$W_y\gg W_x$ and $W_y\to W_x$, respectively) is concomitant with a monotonous
change from HH to LH character for the lowest quasi-1D subbands. As recent hole point
contacts~\cite{romain:apl:06} were realised in quantum wells grown in the low-symmetry
[113] direction,  we also present results for quasi-1D hole states calculated from the full
Luttinger Hamiltonian, including axial and cubic corrections.

\section{2D-to-1D crossover in the spherical approximation}
\begin{SCfigure}[4][t]
\includegraphics[width=.5\textwidth]{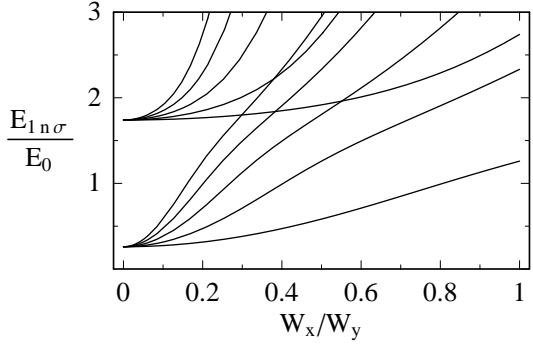}
\caption{Subband energies for hole wires formed by a hard-wall confinement in the $xy$
plane, characterised by widths $W_x$ and $W_y$. We show results for the five lowest
wire subband edges deriving from the lowest 2D HH and LH subbands, respectively.
$E_0 = \gamma_1 \pi^2 \hbar^2/(2 m_0 W_x^2)$, where $\gamma_1$ denotes a
Luttinger parameter~\cite{luttham2} and $m_0$ is the electron mass in vacuum.
The bulk valence-band bottom is taken as zero energy, and $\gamma_{\text{s}}/
\gamma_1=0.37$ was assumed (the value for GaAs~\cite{vurg:jap:01}). We neglected
inter-band HH-LH coupling terms that will change subband crossings into
anticrossings but do not affect wire subbands at energies well below the 2D LH level.} \label{fig:1}
\end{SCfigure}

In the spherical approximation for the kinetic energy, the Hamiltonian for a hole quantum
wire can be written as
${\mathcal H}_{\text{s}}^{\text{(wire)}}={\mathcal H}_{\text{s}}^{\text{(sb)}} + {\mathcal
H}_{\text{s}}^{\text{(hl)}}+{\mathcal H}_{\text{s}}^{\text{(1D)}}  + {\mathcal 
H}_{\text{s}}^{\text{(mix)}}$. These terms correspond to quantised 1D
bound-state energies (sb), a coupling between heavy and light holes arising from
asymmetric 1D confinement (hl), the quadratic energy dispersion for hole motion along
the wire (1D), and a term that accounts for additional inter-subband mixing between
heavy and light holes (mix). Introducing total-angular-momentum ladder operators ${\hat J}_\pm = (\hat J_x \pm i {\hat J}_y)/
\sqrt{2}$, combinations $\hat k_\pm = \hat k_x \pm i \hat  k_y$, and the notation $\{\, , \,
\}$ for a symmetrised product of two operators, the explicit expressions are
\begin{subequations}\label{spheric1D}
\begin{eqnarray}\label{sbsph}
{\mathcal H}_{\text{s}}^{\text{(sb)}} &=& \left\langle \left(\frac{\gamma_1}{2} + 
\frac{\gamma_{\text{s}}}{2} \left[{\hat J}_z^2 - \frac{5}{4}\right]\right) {\hat k}_\perp^2 + 
V(x,y) \right\rangle \quad , \\ \label{hlsph}
{\mathcal H}_{\text{s}}^{\text{(hl)}} &=& -\frac{\gamma_{\text{s}}}{2}\left\langle {\hat k}_x^2
 - {\hat k}_y^2 \right\rangle ({\hat J}_+^2 + {\hat J}_-^2 ) \quad , \\
{\mathcal H}_{\text{s}}^{\text{(1D)}} &=& \left(\frac{\gamma_1}{2} - \gamma_{\text{s}} 
\left[{\hat J}_z^2 - \frac{5}{4}\right]\right) {\hat k}_z^2 \quad , \\
{\mathcal  H}_{\text{s}}^{\text{(mix)}} &=& -\gamma_{\text{s}} \left(\frac{\big\langle \{\hat 
k_x\, , \hat k_y\} \big\rangle}{2i} ( {\hat J}_+^2 - {\hat J}_-^2 ) + \sqrt{2} \left[ \big\langle
\{\hat k_z\, , \hat k_-\} \big\rangle \{\hat J_z\, , \hat J_+\} + \big\langle \{\hat k_z\, , \hat
k_+\} \big\rangle \{\hat J_z\, , \hat J_-\} \right]\right) . \nonumber \\ &&
\end{eqnarray}
\end{subequations}
It is understood that, in Eqs.~(\ref{spheric1D}), operators inside angular brackets have
to be replaced by their corresponding matrix elements between the quantised levels
that diagonalise ${\mathcal H}_{\text{s}}^{\text{(sb)}}$. The term 
${\mathcal H}_{\text{s}}^{\text{(hl)}}$ mixes HH and LH states from the same and also
between different quantised levels of ${\mathcal H}_{\text{s}}^{\text{(sb)}}$. As we are
only interested in hole-wire subbands at energies well below the 2D LH subband edge,
we can neglect the HH-LH mixing between different quantised levels, in particular also
those arising from ${\mathcal H}_{\text{s}}^{\text{(mix)}}$. Quasi-1D hole-subband
energies obtained within this version of the subband $\mathbf{k\cdot p}$ method~\cite
{broido} are shown in Figure~\ref{fig:1}. We obtained analytical expressions, which we
omit here because of space limitations.

\begin{SCfigure}[4][b]
\includegraphics[width=.51\textwidth]{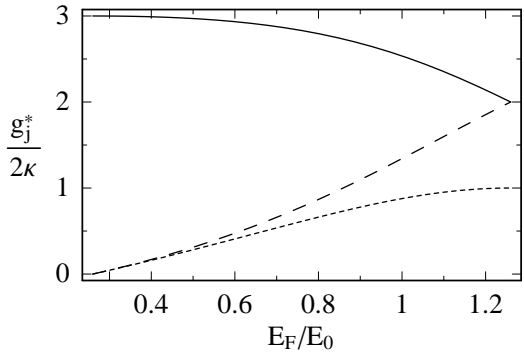}
\caption{Land\'e $g$-factors of hole-wire subband edges crossing the Fermi energy 
$E_{\text{F}}$. The solid (dashed, dotted) curve corresponds to the situation with
magnetic field applied in the $x$ ($y$, $z$) direction. (The wire is aligned with the $z$
axis and the 2D quantum well is grown in the $x$ direction.) The results shown are
obtained using GaAs bandstructure parameters~\cite{vurg:jap:01} and are valid for
magnetic fields that are much smaller than the scale $\tilde B=\pi^2\hbar/(e W_x^2)$.
The Fermi energy is measured from the bulk valence-band bottom. The evolution of the
$g$-factor anisotropy indicates the crossover of hole-wire subbands having HH character
at small $E_{\text{F}}$ (i.e., weak lateral confinement) to LH character in a symmetrically
confined wire.} \label{fig:2}
\end{SCfigure}

In a quantum point contact~\cite{hetero2}, a conductance step occurs whenever a
quasi-1D subband edge passes through the Fermi level $E_{\text{F}}$. The spectroscopic
determination of Zeeman splitting~\cite{patel:prb:91} is based on an analysis of how these
steps change with an applied magnetic field. Hence it is useful to study 1D
hole-subband edges. For comparison with experiments, we provide an approximate
formula for $E_{\text{F}}$ in a 2D HH system with density $n_{\text{2D}}$ and
quantum-well width $W_x$. Measured from the bulk valence-band edge, it is
\begin{equation}
\frac{E_{\text{F}}}{E_0} =  1 - 2 \frac{\gamma_{\text{s}}}{\gamma_1} + \frac{2}{\pi} \left( 1
+ \frac{\gamma_{\text{s}}}{\gamma_1} \right) n_{\text{2D}} W_x^2 \quad .
\end{equation}
Assuming $n_{\text{2D}}=1\times 10^{15}$~m$^{-2}$ and $W_x=20$~nm, which
corresponds to the experimental situation of Ref.~\cite{romain:apl:06}, we
find $E_{\text{F}}/E_0=0.61$. Using Fig.~\ref{fig:1}, we can estimate $W_y\approx 2
W_x$ at the last conductance step.

We have determined the anisotropic hole spin splitting due to an external magnetic field
$\mathbf B$ by diagonalising the sum of the Zeeman term ${\mathcal H}_{\text{Z}}$ and
the spherical Luttinger Hamiltonian with confinement [Eqs.~(\ref{spheric1D})]. Land\'e
$g$-factors for fields parallel to the $x,y,z$ directions can be extracted from the
corresponding Zeeman spin splittings. We focus here on the states at quasi-1D subband
edges, i.e., for $k_z=0$. It turns out that the $g$-factors are only functions of the subband
energy, i.e., have no explicit dependence on the wire's aspect ratio and quasi-1D-level
quantum number. Hence, within the spherical approximation, the hole $g$-factor is the
same for all quasi-1D subbands when they pass the Fermi energy in a point contact. We
show corresponding values in Figure~\ref{fig:2}. Again, space limitations prevent us from
giving analytical results.

\begin{SCfigure}[4][t]
\includegraphics[width=.5\textwidth]{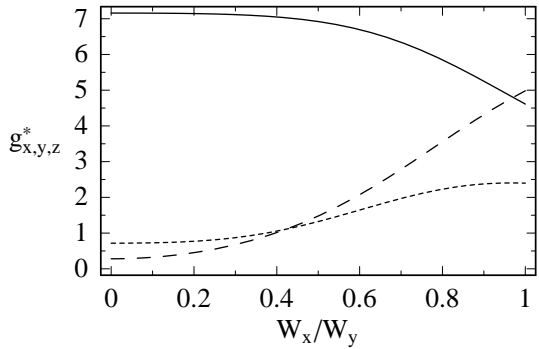}
\caption{Effective $g$-factors as a function of aspect ratio for a hole wire realised in a $[113]$ GaAs quantum well and oriented
parallel to the $[33\bar 2]$ direction. The solid (dashed,
dotted) curve corresponds to the magnetic field being applied in the $x$ ($y$, $z$)
direction. (We consider the geometry where the wire is aligned with the $z$ axis and the
2D quantum well is grown in the $x$ direction.) Results shown are obtained using GaAs
bandstructure parameters~\cite{vurg:jap:01} and are valid for magnetic fields that are
much smaller than the scale $\tilde B=\pi^2 \hbar/(e W_x^2)$ set by the 2D quantum-well
confinement.} \label{fig:3}
\end{SCfigure}

\section{Realistic description of hole wires: Axial and cubic corrections}
The simple spherical model discussed in the previous Section captures the qualitative
2D-to-1D crossover expected in a hole QPC. However, for a proper quantitative
description of the Zeeman splitting, we need to include axial and cubic terms as well. For
example, cubic terms are providing the leading contribution to the $g$-factor for in-plane
field directions in 2D HH systems~\cite{rolandbook} and, hence, can also be expected to
be dominant for wide hole wires. Hence, the dependence of $g$-factors on wire width will
be strongly influenced by the axial and cubic parts.

As before, we find the hole $g$-factors by diagonalising the sum of the Luttinger
Hamiltonian (now including axial and cubic terms) with confinement and the Zeeman
term, again neglecting HH-LH coupling between different quantised orbital levels. Our
results for a quantum-wire geometry realised in recent experiment~\cite{romain:apl:06}
are shown in Figure~\ref{fig:3}. With cubic corrections included, the  $g$-factors obtained
for in-plane magnetic-field directions in the 2D limit ($W_y\gg W_x$) are finite. Their
values agree quite well with those given previously~\cite{rolandbook}, even though our
approach neglects HH-LH mixing between different quantised 1D orbital levels and,
hence, could have been rather bad for describing the 2D limit. Most interestingly, though,
we find that the in-plane $g$-factor anisotropy changes character during the 2D-to-1D
crossover. The crossing of the dashed and dotted curves in Fig.~\ref{fig:3} indicates that
the confinement-induced HH-LH mixing eventually results in a larger Zeeman splitting for
in-plane fields perpendicular to the hole wire as compared with fields parallel to the wire,
effectively reversing the situation encountered in the 2D limit.

\section{Conclusions}
We calculated quasi-1D subbands and $g$-factors for laterally confined 2D HH systems.
Confinement-induced band mixing causes a strongly wire-width-dependent Zeeman
splitting. Future work needs to consider the effect of a nonuniform lateral confinement and 
address many-body effects.

\begin{acknowledgement}
The author thanks O.P.~Sushkov for illuminating discussions and R.~Winkler for a critical
reading of this manuscript. Financial support from the Marsden Fund of the Royal Society
of NZ is gratefully acknowledged. This research was also supported in part by the
National Science Foundation under Grant No.~PHY99-07949 during a visit to the Kavli
Institute for Theoretical Physics at the University of California, Santa Barbara.
\end{acknowledgement}

%\bibliography{general,spintronics,spinorbit,mesophys,myself}

\end{document}